\documentstyle[preprint,aps]{revtex}

\begin{document}
\title{Ground State and Quasiparticle Spectrum of a Two Component Bose-Einstein
Condensate}
\author{C. P. Search, A. G. Rojo, and P. R. Berman}
\address{Michigan Center for Theoretical Physics, Physics Department, University\\
of Michigan, Ann Arbor, MI, 48109-1120.}
\date{\today}
\maketitle

\begin{abstract}
We consider a dilute homogenous atomic Bose-Einstein condensate with two
non-degenerate internal energy levels. We discuss the case in which the two
components achieve a state of chemical equilibrium in the presence of an
external radiation field which couples the two states. The presence of the
radiation field can result in new ground states for the condensate as a
consequence of the lowering of the condensate energy due to the interaction
energy with the field. We analyze the ground state energy as a function of
the coupling constants for the two-body interactions, the Rabi frequency of
the radiation field, and the detuning of the field. We also give explicit
expressions for the quasiparticle excitation spectrum.
\end{abstract}

\section{Introduction}

The recent experimental realization of trapped Bose-Einstein condensates
(BEC) with internal degrees of freedom \cite{exper} has sparked much
theoretical and experimental study of the properties of multicomponent
condensates. Of fundamental importance is the structure of the ground state
and the energy spectrum of collective excitations above the condensate
ground state. This has been explored extensively for the case of spinor
condensates \cite{spinor} in which the internal degrees of freedom
correspond to the different Zeeman states of a particular hyperfine manifold
such as the $F=1$ manifold in $^{23}Na.$ The two-body interaction
Hamiltonian for a spinor condensate is invariant under rotations in spin
space, since for two-body collisions, the s-wave scattering length can
depend only on the total spin angular momentum of the two atoms owing to the
rotational symmetry of the collision. In contrast, for two component
condensates, such as $^{87}Rb,$ in which the two states correspond to
non-degenerate internal states of the atoms (i.e. states which differ in
either their principal quantum number or total angular momentum quantum
numbers), the two-body interaction is not symmetric with respect to
pseudo-spin $SU(2)$ transformations since the s-wave scattering length
depends on the internal eigenstates of the two atoms. If an external field
couples the two internal states, one would expect effects that are not
present in spinor condensates.

The ground state and energy spectrum of the quasiparticle excitations of a
two component homogenous condensate in the presence of a coupling field has
been calculated in Ref. \cite{instable}. The coupling field allows atoms to
make transitions between the two internal states. In that article, as well
as in the calculations to be presented below, it is assumed that the two
components of the condensate are in a state of chemical equilibrium (i. e.
the chemical potential for the two components are equal \cite{kittel}). This
condition is equivalent to assuming that the total number of atoms in the
condensate is fixed rather than the number of atoms in each of the two
components. For nonzero temperature, the relative number of atoms in the two
components is determined by the condition that the atoms are in thermal
equilibrium \cite{griffin}.

However, for zero temperature (which is the case considered in Ref. \cite
{instable}), the chemical potential may be identified with the energy of a
single atom in the condensate. In this case, chemical equilibrium
corresponds to the condensate being in a stationary state of the system. The
stationary states of a free atom interacting with an external radiation
field are known as dressed states in quantum optics \cite{dressed}.
Consequently, chemical equilibrium for a two-component condensate
corresponds to the direct generalization of the atomic dressed states \cite
{dressedconden}.

Under the condition of chemical equilibrium, Goldstein and Meystre \cite
{instable} found that the quasiparticle spectrum could contain states having
imaginary energies, indicating the onset of instabilities in the condensate.
This is a rather surprising and unexpected result; the energy spectrum for
quasiparticle excitations above the ground state of an interacting many body
system are expected to be real. Given these results, we have calculated the
ground state and quasiparticle spectrum for the model of a two component
condensate discussed in \cite{instable}. We obtain an expression for the
ground state energy density as a function of the relative concentration of
the two components of the condensate, given a fixed total number of atoms in
the condensate. For a coupling field of finite strength, the energy density
can exhibit both maxima and minima as a function of the number of atoms in
one of the components. The minima correspond to points of stable
equilibrium. As one might expect, the excitation spectra calculated around
these new minima are real for all momenta. For values of the relative
concentration other than those corresponding to the minima, in the presence
of relaxation, the system will be driven to a point of stable equilibrium.
The instabilities found in Ref. \cite{instable} can be traced to an
expansion about a point of unstable equilibrium.

In section II, the Hamiltonian for a two component condensate, expressed in
a form which emphasizes the broken $SU(2)$ symmetry, is used to study the
ground state energy as a function of the relative fraction of atoms in each
of the two states. In section III, the quasiparticle spectrum is derived.

\section{Hamiltonian and Ground State Energy}

The Hamiltonian operator for a two component BEC may be written as 
\begin{equation}
\hat{H}=\hat{H}_{1}+\hat{H}_{2}  \label{Ho}
\end{equation}
where $\hat{H}_{1}$ is the single body Hamiltonian given by 
\begin{eqnarray}
\hat{H}_{1} &=&\int d^{3}r\left\{ \hat{\Psi}_{a}^{\dagger }({\bf r})\left[ -%
\frac{\hbar ^{2}\nabla ^{2}}{2m}+V_{a}({\bf r})+\frac{\hbar \delta }{2}%
\right] \hat{\Psi}_{a}({\bf r})+\hat{\Psi}_{b}^{\dagger }({\bf r})\left[ -%
\frac{\hbar ^{2}\nabla ^{2}}{2m}+V_{b}({\bf r})-\frac{\hbar \delta }{2}%
\right] \hat{\Psi}_{b}({\bf r})\right.  \nonumber \\
&&\left. +\frac{\hbar {\cal R}}{2}\left[ \hat{\Psi}_{a}^{\dagger }({\bf r})%
\hat{\Psi}_{b}({\bf r})+\hat{\Psi}_{b}^{\dagger }({\bf r})\hat{\Psi}_{a}(%
{\bf r})\right] \right\} ;  \label{Hs}
\end{eqnarray}
and $\hat{H}_{2}$ is the two-body interaction, 
\begin{eqnarray}
\hat{H}_{2} &=&\frac{1}{2}\int d^{3}rd^{3}r^{\prime }\left\{ \hat{\Psi}%
_{a}^{\dagger }({\bf r})\hat{\Psi}_{a}^{\dagger }({\bf r}^{\prime })U_{a}(%
{\bf r-r}^{\prime })\hat{\Psi}_{a}({\bf r}^{\prime })\hat{\Psi}_{a}({\bf r})+%
\hat{\Psi}_{b}^{\dagger }({\bf r})\hat{\Psi}_{b}^{\dagger }({\bf r}^{\prime
})U_{b}({\bf r-r}^{\prime })\hat{\Psi}_{b}({\bf r}^{\prime })\hat{\Psi}_{b}(%
{\bf r})+\right.  \label{Ht} \\
&&\left. 2\hat{\Psi}_{a}^{\dagger }({\bf r})\hat{\Psi}_{b}^{\dagger }({\bf r}%
^{\prime })U_{x}({\bf r-r}^{\prime })\hat{\Psi}_{b}({\bf r}^{\prime })\hat{%
\Psi}_{a}({\bf r})\right\} .
\end{eqnarray}

The operators $\hat{\Psi}_{i}({\bf r})$ are the annihilation operators for
an atom in state $i=\{a,b\}$ at position ${\bf r}$ which satisfy the Bosonic
commutation relation, $\left[ \hat{\Psi}_{i}({\bf r}),\hat{\Psi}%
_{j}^{\dagger }({\bf r}^{\prime })\right] =\delta _{ij}\delta ({\bf r-r}%
^{\prime })$ with all other commutators being zero. There is a spatially
uniform field which couples the two states and has a Rabi frequency given by 
${\cal R}>0$. The atomic field operators have been written in a field
interaction representation which is rotating at the frequency of the
external field, $\omega _{e}.$ Consequently, there appears in Eq. (\ref{Hs})
the detuning, $\delta =\omega _{o}-\omega _{e},$ where $\hbar \omega _{o}$
is the energy difference between the two internal states. In Eq. (\ref{Hs}),
the $V_{i}({\bf r})$ are external potentials which we take to be zero since
we wish to consider only the case of a homogenous condensate.

For the two-body interaction, we use a contact potential, $U_{i}({\bf r-r}%
^{\prime })=V_{i}\delta ({\bf r-r}^{\prime }).$ The coupling constants, $%
V_{i}$, are expressed in terms of the s-wave scattering lengths by $V_{i}=%
\frac{4\pi \hbar ^{2}a_{i}}{m}$ where $a_{a}$ and $a_{b}$ are the scattering
lengths for collisions between two atoms in states $a$ and $b$, respectively
while $a_{x}$ is the scattering length for collisions between atoms in
different internal states. In accordance with \cite{instable}, $\hat{H}_{2}$
can be simplified by assuming that $V_{a}\approx V_{b}=V_{s}\neq V_{x}.$

It is possible to rewrite $\hat{H}_{2}$ in a simplified form that emphasizes
the lack of $SU(2)$ symmetry. To do this one expresses the field operators
as two-component spinors, 
\begin{equation}
\hat{\Psi}({\bf r})=\left( 
\begin{array}{c}
\hat{\Psi}_{a}({\bf r}) \\ 
\hat{\Psi}_{b}({\bf r})
\end{array}
\right) .
\end{equation}
The two-body Hamiltonian may then be expressed as

\begin{mathletters}
\begin{eqnarray}
\hat{H}_{2} &=&\frac{1}{2}\int d^{3}rd^{3}r^{\prime }\hat{\Psi}^{\dagger }(%
{\bf r})\hat{\Psi}^{\dagger }({\bf r}^{\prime }){\bf V}_{I}({\bf r,r}%
^{\prime })\hat{\Psi}({\bf r}^{\prime })\hat{\Psi}({\bf r}); \\
{\bf V}_{I}({\bf r,r}^{\prime }) &=&\frac{1}{2}(V_{s}+V_{x}){\bf I(r)I(r}%
^{\prime }{\bf )}\delta ({\bf r-r}^{\prime })+\frac{1}{2}(V_{s}-V_{x})\delta
({\bf r-r}^{\prime }){\bf \sigma }_{z}({\bf r)\sigma }_{z}({\bf r}^{\prime }%
{\bf );}
\end{eqnarray}
where ${\bf I(r)}$ is the identity matrix and ${\bf \sigma }_{z}({\bf r)}$
is the Pauli spin matrix which act on the spinor at ${\bf r.}$ The single
body Hamiltonian may also be expressed in the compact form, 
\end{mathletters}
\begin{equation}
\hat{H}_{1}=\int d^{3}r\hat{\Psi}^{\dagger }({\bf r})\left\{ -\frac{\hbar
^{2}\nabla ^{2}}{2m}{\bf I}+\frac{\hbar \delta }{2}{\bf \sigma }_{z}{\bf +}%
\frac{\hbar {\cal R}}{2}{\bf \sigma }_{x}\right\} \hat{\Psi}({\bf r}).
\end{equation}
The previously mentioned lack of $SU(2)$ symmetry in $\hat{H}_{2}$ is now
obvious. If we carry out an active rotation that diagonalizes $\hat{H}_{1}$, 
$\hat{H}_{2}$ will not be invariant since the interactions are proportional
to ${\bf \sigma }_{z}({\bf r)\sigma }_{z}({\bf r}^{\prime }{\bf )}$ instead
of ${\bf \sigma }({\bf r)\cdot \sigma }({\bf r}^{\prime }{\bf ).}$

One can now write a c-number function for the ground state energy. This is
accomplished by rewriting the field operators as $\hat{\Psi}_{i}({\bf r}%
)=\phi _{i}({\bf r})+\delta \hat{\Psi}_{i}({\bf r})$ where $\phi _{i}({\bf r}%
)$ is the condensate wave function which is defined as $\phi _{i}({\bf r}%
)\equiv \left\langle \hat{\Psi}_{i}({\bf r})\right\rangle =\sqrt{\rho _{i}}%
e^{iS_{i}}$ and the expectation value is taken with respect to the
condensate state \cite{lifshitz}. Here $\rho ({\bf r})=\rho _{a}({\bf r}%
)+\rho _{b}({\bf r})$ is the total density which is conserved and $%
S_{ab}=S_{a}-S_{b}$ is the relative phase between the two components. By
making the replacement $\hat{\Psi}_{i}({\bf r})\rightarrow \phi _{i}({\bf r})
$ in Eq. (\ref{Ho}), one obtains an energy functional for the condensate.
For a homogenous stationary system, Bose-Einstein condensation occurs in the
state with zero momentum and as such, the $\phi _{i}({\bf r})$ are
independent of ${\bf r}$. The energy density for the homogenous condensate
with volume $V$ is, 
\begin{mathletters}
\begin{eqnarray}
\frac{E_{o}}{V} &=&\frac{\hbar \delta }{2}\left( \rho _{a}-\rho _{b}\right)
+\hbar {\cal R}\sqrt{\rho _{a}\rho _{b}}\cos (S_{ab})+\frac{1}{4}%
(V_{s}-V_{x})\left( \rho _{a}-\rho _{b}\right) ^{2}+\frac{1}{4}%
(V_{s}+V_{x})\rho ^{2}  \label{ground} \\
&=&\frac{\hbar \delta }{2}\left( 2\rho _{a}-\rho \right) +\hbar {\cal R}%
\sqrt{\rho _{a}(\rho -\rho _{a})}\cos (S_{ab})+\frac{1}{4}%
(V_{s}-V_{x})\left( 2\rho _{a}-\rho \right) ^{2}+\frac{1}{4}%
(V_{s}+V_{x})\rho ^{2}
\end{eqnarray}
Notice that Eq. (\ref{ground}) may also be derived from $\left\langle \hat{H}%
\right\rangle =E_{o}-\frac{1}{2\rho }\left( V_{s}\left( \rho _{a}^{2}+\rho
_{b}^{2}\right) +2V_{x}\rho _{a}\rho _{b}\right) $ where the expectation
value is taken with respect to the wave function 
\end{mathletters}
\begin{equation}
\left| C\right\rangle =\frac{1}{\sqrt{N_{o}!}}\left( \sqrt{\frac{N_{oa}}{%
N_{o}}}e^{iS_{a}}\hat{a}_{a{\bf 0}}^{\dagger }+\sqrt{\frac{N_{ob}}{N_{o}}}%
e^{iS_{b}}\hat{a}_{b{\bf 0}}^{\dagger }\right) ^{N_{o}}\left| 0\right\rangle 
\label{Ncons}
\end{equation}
where $\hat{a}_{i{\bf 0}}^{\dagger }$ is the creation operator for an atom
in state i with zero momentum (see next section), $N_{oi}=\rho _{i}V$, and $%
N_{o}=N_{oa}+N_{ob}$. Note that $\frac{1}{2\rho }\left( V_{s}\left( \rho
_{a}^{2}+\rho _{b}^{2}\right) +2V_{x}\rho _{a}\rho _{b}\right) $ is an
intensive quantity whereas the total energy, $E_{o},$ is extensive and
consequently $\frac{1}{2\rho }\left( V_{s}\left( \rho _{a}^{2}+\rho
_{b}^{2}\right) +2V_{x}\rho _{a}\rho _{b}\right) $ is negligible in the
thermodynamic limit. Finally, it is clear from Eq. (\ref{Ncons}) that $%
S_{ab}=0$ and $\pi $ correspond to symmetric and antisymmetric
superpositions of the internal atomic states.

In the remainder of the paper, we consider the case in which there is a
nonvanishing coupling between the two components which allows the atoms to
make transitions between states $a$ and $b$. In other words, ${\cal R}$ is
never identically zero although it may be infinitesimally small, ${\cal R}%
\sim 0.$ If ${\cal R}\equiv 0$, then the number of atoms in states $a$ and $b
$ would be separately conserved and the condensate energy would be given Eq.
(\ref{ground}) for fixed $\rho _{a}$ and $\rho _{b}.$ However, when
Bose-Einstein condensation occurs in the presence of a coupling fied, the
condensate will form in that state which minimizes Eq. (\ref{ground}) for
fixed total number of atoms ($\rho =$const.); this is the limit we consider.
For non-interacting atoms, this would correspond to the lowest energy
dressed state.

Consequently, the ground state of the condensate will be a function of $\rho
,$ $\delta ,$ ${\cal R}$, $S_{ab}$, and $V_{s}-V_{x}$. Note that Steel and
Collett have carried out a fully quantum mechanical calculation of the
ground state for small condensates (a few hundred atoms) under the same
conditions \cite{steel}. Similarly, the ground state of a one dimensional
inhomogenous condensate has been studied under these conditions by Blakie 
{\it et al.} using numerical solutions of the time independent coupled
Gross-Pitaevskii equations \cite{dressedconden}. In both these cases,
however, the relative roles played by the mean-field interactions and
interaction with the external field in determining the ground state of the
condensate is not as physically clear as it is for the case of a homogenous
condensate.

In the following two subsections we classify the extrema of Eq. (\ref{ground}%
) for the two cases of ${\cal R}\sim 0$ and $\delta =0.$ The key point is
that nonzero $\left( V_{s}-V_{x}\right) $ can significantly modify the
ground state structure from what one would expect based on the single body
Hamiltonian. However, before proceeding it is helpful to make a few
definitions. We define the polarization of the condensate to be $\xi =\left|
\rho _{a}-\rho _{b}\right| /\rho $. Consequently, a polarized condensate
corresponds to $\xi =1$ and an unpolarized condensate would correspond to $%
\xi =0$ while $0<\xi <1$ represents a state of partial polarization.

\subsection{${\cal R}\sim 0$}

In this case we set ${\cal R}$ equal to zero in Eq. (\ref{ground}). When $%
\delta =0$, the ground state will either be unpolarized for $V_{s}>V_{x}$ or
polarized for $V_{s}<V_{x}.$ When $\delta >0$ and $V_{s}<V_{x},$ the ground
state corresponding to the minimum of $E_{o}/V$ occurs at $\rho _{b}=\rho $,
while $E_{o}/V$ is a maximum at $\rho _{a}=\frac{-\hbar \delta }{%
2(V_{s}-V_{x})}+\rho /2.$ For the case $V_{s}>V_{x}$ (again with $\delta >0$%
), $\rho _{a}=\frac{-\hbar \delta }{2(V_{s}-V_{x})}+\rho /2$ now corresponds
to the minimum for $E_{o}/V.$ Notice that when ${\cal R}\sim 0$, the
relative phase, $S_{ab}$, is arbitrary. For $\delta <0$, the results are the
same if one interchanges states $a$ and $b$. Note that for binary
condensates, where $N_{a}$ and $N_{b}$ are seperately conserved, $%
V_{s}<V_{x} $ would lead to a phase separation of the condensates into two
components that occupy nonoverlapping regions in space \cite{ho-shenoy}\cite
{binary} \cite{phase}.

\subsection{$\protect\delta =0$}

When ${\cal R}$ is finite, the ground state can exhibit interesting new
structure. It is easy to see from Eq. (\ref{ground}), that the energy is an
extremum only if $S_{ab}=0,\pi .$ By requiring that $E_{o}/V$ be an extremum
with respect to $\rho _{a},$ one finds from $\partial \left( E_{o}/V\right)
/\partial \rho _{a}$, that the extrema are located at $\rho _{a}=\rho /2$
and $\rho _{a}=\rho _{a}^{(\pm )}$ where 
\begin{equation}
\rho _{a}^{(\pm )}\equiv \frac{\rho }{2}\left( 1\pm \sqrt{1-1/(2\rho \alpha
)^{2}}\right)
\end{equation}
and 
\begin{equation}
\alpha =\frac{V_{s}-V_{x}}{2\hbar {\cal R}}.
\end{equation}
Note that the extrema at $\rho _{a}^{(\pm )}$ are degenerate and occur only
when $\left| \rho \alpha \right| >1/2$ and $\alpha /\cos (S_{ab})>0.$ There
are four cases to consider:

{\it (i) }$\rho \alpha <1/2${\it \ and }$S_{ab}=0.$ In this case the only
extremum is at $\rho _{a}=\rho /2$ and this is a global maximum of $E_{o}/V.$

{\it (ii) }$\rho \alpha >1/2${\it \ and }$S_{ab}=0.$ In this case $\rho
_{a}=\rho /2$ is now a global minimum of $E_{o}/V.$ The extrema at $\rho
_{a}=\rho _{a}^{(\pm )}$ are global maxima of $E_{o}/V$.

{\it (iii) }$\rho \alpha >-1/2${\it \ and }$S_{ab}=\pi .$ Again, $\rho
_{a}=\rho /2$ is a global minimum of $E_{o}/V.$

{\it (iv)\ }$\rho \alpha <-1/2${\it \ and }$S_{ab}=\pi .$ The extremum at $%
\rho _{a}=\rho /2$ has now become the global maximum of $E_{o}/V$. The
extrema at $\rho _{a}=\rho _{a}^{(\pm )}$ are the global minima in the
ground state energy. Note that when $\rho \alpha \rightarrow \infty $, the
minima are located at $\rho _{a}=0$ or $\rho _{a}=\rho $.

The nature of these extrema may be understood by a consideration of the
physical meaning of the parameter $\rho \alpha .$ Since the $\left(
V_{s}+V_{x}\right) $ term in $E_{o}/V$ simply gives an overall constant
energy, it may neglected and, as such, the only relevant mean field
interaction energy for determining the ground state is the $\frac{1}{4}%
(V_{s}-V_{x})\left( \rho _{a}-\rho _{b}\right) ^{2}$ term. Therefore, $\rho
\alpha $ is the ratio of the mean-field energy per atom, $\sim
(V_{s}-V_{x})\rho $, to the atom-field interaction energy, $\sim \hbar {\cal %
R}$. Consequently, for $\left| \rho \alpha \right| >1/2$ the mean-field
interactions dominate the ground state energy while for $\left| \rho \alpha
\right| <1/2$ the energy of the ground state is dominated by the atom-field
interaction energy. For $\rho \alpha <-1/2$ the extremum at $\rho _{a}=\rho
/2$ is a maximum, regardless of $S_{ab}$, since the mean-field interactions
favor a polarized ground state ($\xi =1$) in this limit. Similarly, for $%
\rho \alpha >1/2,$ $\rho _{a}=\rho /2$ is always a minimum since the
mean-field interactions favor an unpolarized state ($\xi =0$).

On the other hand, for $-1/2<\rho \alpha <1/2,$ the single body atom-field
interaction dominates over the mean-field interactions. In this limit, the
minimum of $E_{o}/N=\rho ^{-1}(E_{o}/V)$ coincides with the lowest energy
dressed state of the free atoms. For $\delta =0,$ the lowest energy dressed
is the antisymmetric state $\frac{1}{\sqrt{2}}\left( \left| a\right\rangle
-\left| b\right\rangle \right) $ with energy $-\hbar {\cal R}/2$ which
corresponds to $\rho _{a}=\rho /2$ and $S_{ab}=\pi $. Consequently, the
state $\rho _{a}=\rho /2$ and $S_{ab}=\pi $ is the global minimum of $%
E_{o}/V $ in the interval $-1/2<\rho \alpha <1/2$. The other dressed state
for $\delta =0$ is the symmetric state $\frac{1}{\sqrt{2}}\left( \left|
a\right\rangle +\left| b\right\rangle \right) $ with energy $\hbar {\cal R}%
/2 $ which is not the ground state of the system. The symmetric state
corresponds to $\rho _{a}=\rho /2$ and $S_{ab}=0$ which is a global maximum
in the interval $-1/2<\rho \alpha <1/2.$

For case (ii) and (iv), the energy difference between the extremum at $\rho
/2$ and $\rho _{a}^{(+)}$ is given by 
\begin{equation}
\Delta E/V=\frac{1}{V}\left( E(\rho _{a}=\rho /2)-E(\rho _{a}=\rho
_{a}^{(+)})\right) =\frac{\rho \hbar {\cal R}}{2}\left[ -\frac{1}{4\rho
\alpha }-\rho \alpha +\cos (S_{ab})\right] .
\end{equation}
The energy difference helps to elucidate the transition of the central
extremum at $\left| \rho \alpha \right| =1/2.$ One can see that as $\left|
\rho \alpha \right| \rightarrow 1/2$ from above, $\rho _{a}^{(\pm
)}\rightarrow \rho /2$ and $\Delta E/V\rightarrow 0$ so that the three
extrema merge at $\left| \rho \alpha \right| =1/2$ and for $\left| \rho
\alpha \right| <1/2$, there is a single extremum at $\rho _{a}=\rho /2.$
This is illustrated in Figures 1 and 2.

One should note that $E_{o}/V$ always possess at least one global minimum
for finite ${\cal R}$ but this minimum does not necessarily correspond to
the state with equal population in the two components \cite{instable}. It is
interesting to note that for $\rho \alpha <-1/2$ the condensate state
exhibits another broken symmetry in addition to the usual broken $U(1)$
gauge symmetry since condensation occurs at either $\rho _{a}^{(+)}$ or $%
\rho _{a}^{(-)}$.

\section{Quasiparticle Spectrum}

In this section the Bogoliubov prescription is used to linearize the
Hamiltonian around the ground states discussed in the previous section. A
canonical transformation is then used to diagonalize the Hamiltonian and
find the spectrum of elementary excitations above the condensate.

At this point it is advantageous to introduce the grand canonical
Hamiltonian, $\hat{K}=\hat{H}-\mu \hat{N}$. Here, $\hat{N}=\hat{N}_{a}+\hat{N%
}_{b}=\int d^{3}r\left( \hat{\Psi}_{a}^{\dagger }({\bf r})\hat{\Psi}_{a}(%
{\bf r})+\hat{\Psi}_{b}^{\dagger }({\bf r})\hat{\Psi}_{b}({\bf r})\right) $
is the total number operator and $\mu $ is the chemical potential. We have
assumed that the system is in a state of chemical equilibrium so that $\mu
=\mu _{a}=\mu _{b}$ where $\mu _{i}=\frac{\partial E}{\partial N_{i}}$ is
the chemical potential of the two components. The chemical potential insures
that the ground state expectation value, $\left\langle \hat{K}\right\rangle $%
, is a minimum with respect to the total number of atoms. The chemical
potential may also be interpreted as a Lagrange multiplier which insures
that $\left\langle \hat{N}\right\rangle $ is conserved \cite{fetter}. A pair
of equations for $\mu $ may be derived by requiring that $E_{o}[\phi
_{a},\phi _{b}]-\mu \int d^{3}r\left( \left| \phi _{a}\right| ^{2}+\left|
\phi _{b}\right| ^{2}\right) $ be an extremum with respect to the variations 
$\delta \phi _{a}^{\ast }$ and $\delta \phi _{b}^{\ast },$%
\begin{mathletters}
\begin{eqnarray}
\mu \phi _{a} &=&\frac{\hbar \delta }{2}\phi _{a}+\frac{\hbar {\cal R}}{2}%
\phi _{b}+\left( V_{s}\left| \phi _{a}\right| ^{2}+V_{x}\left| \phi
_{b}\right| ^{2}\right) \phi _{a};  \label{chem1} \\
\mu \phi _{b} &=&-\frac{\hbar \delta }{2}\phi _{b}+\frac{\hbar {\cal R}}{2}%
\phi _{a}+\left( V_{s}\left| \phi _{b}\right| ^{2}+V_{x}\left| \phi
_{a}\right| ^{2}\right) \phi _{b}.  \label{chem2}
\end{eqnarray}
It should be noted that Eqs. (\ref{chem1}-\ref{chem2}) are equivalent to the
condition that $\left\langle \hat{K}\right\rangle $ is an extremum with
respect to the {\it total }number of atoms which is found by varying $N_{oa}$
while keeping $N_{ob}$ fixed and vice versa. In contrast, the extrema for $%
E_{o}$ found in section II corresponded to finding $\left( \frac{\partial
E_{o}}{\partial N_{oa}}\right) _{N_{o}}=0$, i.e. the extrema for variations
in $N_{oa}$ for a fixed total number of atoms in the condensate. As such,
Eqs. (\ref{chem1}- \ref{chem2}) serve to define the chemical potential, but 
{\it not} the relative fraction of $\phi _{a}$ and $\phi _{b},$ which is
determined by minimizing Eq. (\ref{ground}).

For a homogenous condensate we can expand $\hat{\Psi}_{i}({\bf r})$ in a
basis of plane wave states, $\hat{\Psi}_{i}({\bf r})=\frac{1}{\sqrt{V}}\sum_{%
{\bf p}}\hat{a}_{i{\bf p}}e^{i{\bf p\cdot r}/\hbar }$, with $\phi _{i}=\frac{%
1}{\sqrt{V}}\left\langle \hat{a}_{i{\bf 0}}\right\rangle $ and $\left[ \hat{a%
}_{i{\bf p}},\hat{a}_{j{\bf p}^{\prime }}^{\dagger }\right] =\delta
_{ij}\delta _{{\bf p},{\bf p}^{\prime }}$. This gives the following
expression for $\hat{K},$%
\end{mathletters}
\begin{eqnarray}
\hat{K} &=&\sum_{{\bf p}}\left\{ \left( \frac{p^{2}}{2m}+\frac{\hbar \delta 
}{2}-\mu \right) \hat{a}_{a{\bf p}}^{\dagger }\hat{a}_{a{\bf p}}+\left( 
\frac{p^{2}}{2m}-\frac{\hbar \delta }{2}-\mu \right) \hat{a}_{b{\bf p}%
}^{\dagger }\hat{a}_{b{\bf p}}+\frac{\hbar {\cal R}}{2}\left( \hat{a}_{a{\bf %
p}}^{\dagger }\hat{a}_{b{\bf p}}+\hat{a}_{b{\bf p}}^{\dagger }\hat{a}_{a{\bf %
p}}\right) \right\}  \nonumber \\
&&+\frac{1}{2V}\sum_{{\bf p}_{1}+{\bf p}_{2}={\bf p}_{3}+{\bf p}_{4}}\left\{
V_{s}\left( \hat{a}_{a{\bf p}_{1}}^{\dagger }\hat{a}_{a{\bf p}_{2}}^{\dagger
}\hat{a}_{a{\bf p}_{3}}\hat{a}_{a{\bf p}_{4}}+\hat{a}_{b{\bf p}%
_{1}}^{\dagger }\hat{a}_{b{\bf p}_{2}}^{\dagger }\hat{a}_{b{\bf p}_{3}}\hat{a%
}_{b{\bf p}_{4}}\right) +2V_{x}\hat{a}_{a{\bf p}_{1}}^{\dagger }\hat{a}_{b%
{\bf p}_{2}}^{\dagger }\hat{a}_{a{\bf p}_{3}}\hat{a}_{b{\bf p}_{4}}\right\} ;
\end{eqnarray}
which may be linearized around the ground state solutions found in section
II by making the replacement $\hat{a}_{i{\bf 0}}\rightarrow \left\langle 
\hat{a}_{i{\bf 0}}\right\rangle $ and keeping only the lowest order
quadratic terms in the operators for ${\bf p}\neq 0.$ By utilizing Eqs. (\ref
{chem1}-\ref{chem2}), the resulting expression for $\hat{K}$ is 
\begin{eqnarray}
\hat{K} &=&E_{o}-\mu N_{o}+\sum_{{\bf p\neq 0}}\left\{ \left( \frac{p^{2}}{2m%
}+\rho _{a}V_{s}-\frac{\hbar {\cal R}}{2}\sqrt{\frac{\rho _{b}}{\rho _{a}}}%
\cos S_{ab}\right) \hat{a}_{a{\bf p}}^{\dagger }\hat{a}_{a{\bf p}}+\left( 
\frac{p^{2}}{2m}+\rho _{b}V_{s}-\frac{\hbar {\cal R}}{2}\sqrt{\frac{\rho _{a}%
}{\rho _{b}}}\cos S_{ab}\right) \hat{a}_{b{\bf p}}^{\dagger }\hat{a}_{b{\bf p%
}}\right.  \nonumber \\
&&\left. +\left( \frac{\hbar {\cal R}}{2}+V_{x}\sqrt{\rho _{a}\rho _{b}}\cos
S_{ab}\right) \left( \hat{a}_{a{\bf p}}^{\dagger }\hat{a}_{b{\bf p}}+\hat{a}%
_{b{\bf p}}^{\dagger }\hat{a}_{a{\bf p}}\right) +\frac{1}{2}\rho
_{a}V_{s}\left( \hat{a}_{a{\bf p}}^{\dagger }\hat{a}_{a-{\bf p}}^{\dagger }+%
\hat{a}_{a{\bf p}}\hat{a}_{a-{\bf p}}\right) \right.  \nonumber \\
&&\left. +\frac{1}{2}\rho _{b}V_{s}\left( \hat{a}_{b{\bf p}}^{\dagger }\hat{a%
}_{b-{\bf p}}^{\dagger }+\hat{a}_{b{\bf p}}\hat{a}_{b-{\bf p}}\right) +V_{x}%
\sqrt{\rho _{a}\rho _{b}}\cos S_{ab}\left( \hat{a}_{a{\bf p}}^{\dagger }\hat{%
a}_{b-{\bf p}}^{\dagger }+\hat{a}_{a{\bf p}}\hat{a}_{b-{\bf p}}\right)
\right\} .  \label{Klin}
\end{eqnarray}

A Hamiltonian which is quadratic in bosonic operators with the general form, 
\begin{equation}
\hat{H}=H_{o}+\frac{1}{2}\sum_{i,j,{\bf p}\neq {\bf 0}}A_{ij}({\bf p})\hat{a}%
_{i{\bf p}}^{\dagger }\hat{a}_{j-{\bf p}}^{\dagger }+\frac{1}{2}\sum_{i,j,%
{\bf p}\neq {\bf 0}}A_{ij}^{\ast }({\bf p})\hat{a}_{i{\bf p}}\hat{a}_{j-{\bf %
p}}+\sum_{i,j,{\bf p}\neq {\bf 0}}B_{ij}({\bf p})\hat{a}_{i{\bf p}}^{\dagger
}\hat{a}_{j{\bf p}}.  \label{quadH}
\end{equation}
where $B_{ij}({\bf p})$ is Hermitian and $A_{ij}({\bf p})$ is a symmetric
matrix may be diagonalized by a canonical transformation. This is done by
defining quasiparticle annihilation and creation operators, $\hat{b}_{l{\bf p%
}}$ and $\hat{b}_{l{\bf p}}^{\dagger }$, respectively, given by 
\begin{equation}
\hat{b}_{l{\bf p}}=\sum_{i}\left( u_{li}^{\ast }({\bf p})\hat{a}_{i{\bf p}}%
{\bf -}v_{li}^{\ast }({\bf p})\hat{a}_{i-{\bf p}}^{\dagger }\right)
\label{quasi}
\end{equation}
which satisfy bosonic commutation relations. The quasiparticle operators
obey the equation of motion 
\begin{equation}
\left[ \hat{b}_{l{\bf p}},\hat{H}\right] =\varepsilon _{l}({\bf p})\hat{b}_{l%
{\bf p}};  \label{heisen}
\end{equation}
which is consistent with $\hat{H}=\sum_{l,{\bf p}\neq {\bf 0}}\varepsilon
_{l}({\bf p})\hat{b}_{l{\bf p}}^{\dagger }\hat{b}_{l{\bf p}}+E_{vac}.$ Using
Eqs. (\ref{heisen}) and (\ref{quasi}) it is easy to show that the
quasiparticle energies satisfy the eigenvalue equation \cite{Bogo}, 
\begin{mathletters}
\begin{eqnarray}
\varepsilon _{l}({\bf p})u_{li}({\bf p}) &=&\sum_{j}\left( B_{ij}({\bf p}%
)u_{lj}({\bf p})+A_{ij}({\bf p})v_{lj}({\bf p})\right) ;  \nonumber \\
-\varepsilon _{l}({\bf p})v_{li}({\bf p}) &=&\sum_{j}\left( B_{ij}^{\ast }(%
{\bf p})v_{lj}({\bf p})+A_{ij}^{\ast }({\bf p})u_{lj}({\bf p})\right) ;
\label{eig}
\end{eqnarray}
Since $\varepsilon _{l}({\bf p})$ are not the eigenvalues of a Hermitian
matrix, there is no guarantee that they will be real. However, when the $%
\varepsilon _{l}({\bf p})$ are complex, the $\hat{b}_{l{\bf p}}$ and $\hat{b}%
_{l{\bf p}}^{\dagger }$ do not satisfy bosonic commutation relations and
consequently, the quasiparticles may no longer be interpreted as bosons. In
addition, if $\left( u_{li}({\bf p}),v_{li}({\bf p})\right) $ are a solution
with eigenvalue $\varepsilon _{l}({\bf p}),$ then $\left( v_{li}^{\ast }(%
{\bf p}),u_{li}^{\ast }({\bf p})\right) $ are a solution with eigenvalue $%
-\varepsilon _{l}({\bf p}).$ However, only $\varepsilon _{l}({\bf p})>0$ are
physically significant since the energy of the system must be bounded from
below.

Since Eq. (\ref{Klin}) has the form of Eq. (\ref{quadH}), one may directly
apply Eqs. (\ref{eig}) to calculate the energy spectrum of the quasiparticle
excitations above the condensate. As in the previous section, we focus on
the two cases of ${\cal R}\sim 0$ and $\delta =0$ and limit the discussion
to the case $V_{x}>V_{s}$ ($\alpha <0$).

\subsection{${\cal R}\sim 0$}

For $\delta >0$, the ground state is given by $\rho _{b}=\rho $ and $\rho
_{a}=0.$ In this case Eq. (\ref{Klin}) has the form 
\end{mathletters}
\begin{equation}
\hat{K}=E_{o}-\mu N_{o}+\sum_{{\bf p\neq 0}}\left( \frac{p^{2}}{2m}\hat{a}_{a%
{\bf p}}^{\dagger }\hat{a}_{a{\bf p}}+\varepsilon _{B}(p)\hat{B}_{{\bf p}%
}^{\dagger }\hat{B}_{{\bf p}}\right)  \label{R=0}
\end{equation}
where $\varepsilon _{B}(p)=\sqrt{\frac{p^{2}}{2m}\left( \frac{p^{2}}{2m}%
+2\rho V_{s}\right) }$ and $\hat{B}_{{\bf p}}$ is the quasiparticle
annihilation operator for excitations in state $b$ and is related to $\hat{a}%
_{b{\bf p}}$ through the canonical transformation, $\hat{a}_{b{\bf p}}=\cosh
\varphi _{p}\hat{B}_{{\bf p}}-\sinh \varphi _{p}\hat{B}_{-{\bf p}}^{\dagger
} $ \cite{fetter}. For $\delta <0$, condensation occurs in $\rho _{a}=\rho $
and Eq. (\ref{R=0}) remains valid with the interchange of $\hat{a}_{a{\bf p}%
}\leftrightarrow \hat{a}_{b{\bf p}}$. Finally, for $\delta =0$ and $%
V_{x}>V_{s},$ condensation only occurs in the state $\rho _{b}=\rho $ {\it %
or }$\rho _{a}=\rho $ since the two states are degenerate. Bose condensation
will not occur in the state $\rho _{b}=\rho _{a}=\rho /2$ as given in Ref. 
\cite{instable} since this represents a maximum in $E_{o}/V$.

\subsection{$\protect\delta =0,$ ${\cal R}>0$}

For $\rho \alpha <0$ and $\delta =0$, the only minima in $E_{o}/V$ occur for 
$S_{ab}=\pi .$ For the extremum at $\rho _{a}=\rho /2,$ the quasiparticle
energies are 
\begin{mathletters}
\begin{eqnarray}
\varepsilon _{-}({\bf p}) &=&\sqrt{\frac{p^{2}}{2m}\left( \frac{p^{2}}{2m}%
+\rho \left( V_{s}+V_{x}\right) \right) };  \label{Esym1} \\
\varepsilon _{+}({\bf p}) &=&\sqrt{\left( \frac{p^{2}}{2m}+\hbar {\cal R}%
\right) \left( \frac{p^{2}}{2m}+\hbar {\cal R+}\rho \left(
V_{s}-V_{x}\right) \right) };  \label{Esym2}
\end{eqnarray}
which agrees with the results obtained in Ref. \cite{instable}. It is
possible for $\varepsilon _{+}({\bf p})$ to be imaginary and this situation
corresponds to the instabilities mentioned in \cite{instable}. The condition
that $\varepsilon _{+}({\bf p})$ be real for all momenta is $\rho \alpha
\geq -1/2$ which is exactly the condition that $\rho _{a}=\rho /2$ be a
minimum of $E_{o}/V.$ Consequently, Eqs. (\ref{Esym1}-\ref{Esym2}) can be
interpreted as the quasiparticle spectrum for excitations above the
condensate {\it only} when $\rho \alpha \geq -1/2.$ When Eqs. (\ref{Esym1}-%
\ref{Esym2}) are extended to $\rho \alpha <-1/2$, they no longer correspond
to physically meaningful results since $\varepsilon _{\pm }({\bf p})$ then
correspond to expansions of small amplitude oscillations around an energy
maximum.

When $\rho \alpha <-1/2,$ the location of the minima of $E_{o}/V$ are $\rho
_{a}=\rho _{a}^{(\pm )}$. Since Eq. (\ref{Klin}) is symmetric under the
interchange of $a$ and $b$ atoms, the quasiparticle spectrum will be the
same at $\rho _{a}^{(+)}$ and $\rho _{a}^{(-)}$. The excitation spectrum
about $\rho _{a}^{(-)}$ (or $\rho _{a}^{(+)}$) is significantly more
complicated than at $\rho _{a}=\rho /2$ and is given by 
\end{mathletters}
\begin{mathletters}
\begin{eqnarray}
\varepsilon _{\pm }({\bf p}) &=&\sqrt{\Xi _{1}\pm \frac{1}{2}\sqrt{\Xi _{2}}}%
;  \label{asymquasi} \\
\Xi _{1} &=&\frac{p^{2}}{2m}\left( \frac{p^{2}}{2m}+\rho V_{x}\right) +\frac{%
1}{2}(\hbar {\cal R})^{2}\left( (2\rho \alpha )^{2}-1\right) \\
\Xi _{2} &=&(\hbar {\cal R})^{4}\left( (2\rho \alpha )^{2}-1\right)
^{2}+4(\hbar {\cal R})^{2}\left( (2\rho \alpha )^{2}-1\right) \frac{p^{2}}{2m%
}\left[ -\left( \hbar {\cal R}(2\rho \alpha )+\rho V_{s}\right) +\left( 1+%
\frac{1}{2\rho \alpha }\right) \frac{p^{2}}{2m}\right]  \nonumber \\
&&+4\left( \frac{p^{2}}{2m}\right) ^{2}\left[ \left( \rho V_{s}\right)
^{2}+\left( 1+\frac{1}{2\rho \alpha }\right) (1-2\rho \alpha )(\hbar {\cal R}%
)(\hbar {\cal R}-2\rho V_{s})\right] ;  \label{asymquasi3}
\end{eqnarray}
Note that the $\pm $ in $\varepsilon _{\pm }({\bf p})$ correspond to the two
branches of the excitation spectrum for each of $\rho _{a}^{(+)}$ and $\rho
_{a}^{(-)}$ and not to $\rho _{a}^{(+)}$ and $\rho _{a}^{(-)}$ individually.
It is obvious that $\Xi _{1}\geq 0$ and one can prove that $\Xi _{2}\geq 0$
for $2\rho \alpha \leq -1$. One may also show that $\Xi _{1}^{2}-\Xi
_{2}/4\geq 0$ which implies that $\varepsilon _{\pm }({\bf p})$ will be real
for $2\rho \alpha \leq -1.$ Thus the quasiparticle spectrum is real as
expected.

When ${\bf p}=0$, $\varepsilon _{-}(0)=0$ and $\varepsilon _{+}(0)=\sqrt{%
\rho ^{2}\left( V_{x}-V_{s}\right) ^{2}-(\hbar {\cal R})^{2}}$ which shows
that one branch is gapless while the other branch contains a gap for $\rho
\alpha <-1/2$. It is easy to show that when $\rho \alpha =-1/2$, Eqs. (\ref
{asymquasi}-\ref{asymquasi3}) agree with Eqs. (\ref{Esym1}-\ref{Esym2})
which indicates that the quasiparticle spectrum varies continuously with $%
\rho \alpha .$ There are several important limiting cases for Eq. (\ref
{asymquasi}). First one may consider the limit that $\rho \alpha \rightarrow
-1/2$ while $\frac{p^{2}}{2m},\rho V_{s}\sim \hbar {\cal R}.$ In this case
the excitation spectrum reduces to 
\end{mathletters}
\begin{equation}
\varepsilon _{\pm }({\bf p})=\sqrt{\frac{p^{2}}{2m}\left( \frac{p^{2}}{2m}%
+\rho \left( V_{x}\pm V_{s}\right) \right) }.  \label{nophon}
\end{equation}
Note that Eq. (\ref{nophon}) is not valid in the long wavelength phonon
regime since $\frac{p^{2}}{2m}\sim \hbar {\cal R}\sim \rho \left(
V_{x}-V_{s}\right) .$ Another simple case is when the two minima approach
the edges at $\rho _{a}=0$ and $\rho _{a}=\rho $ which corresponds to the
condition $\left| \rho \alpha \right| \gg 1/2$ while at the same time $\frac{%
p^{2}}{2m}\ll \hbar {\cal R},\rho V_{x},\rho V_{s}$ (long wavelength limit).
In this case the two branches of Eq. (\ref{asymquasi}) are given by 
\begin{mathletters}
\begin{eqnarray}
\varepsilon _{-}({\bf p}) &=&\sqrt{\frac{\rho V_{s}}{m}}p; \\
\varepsilon _{+}({\bf p}) &=&\sqrt{p^{2}\frac{\rho \left( V_{x}-V_{s}\right) 
}{m}+(\hbar {\cal R})^{2}\left( (2\rho \alpha )^{2}-1\right) }; \\
&\approx &\rho \left( V_{x}-V_{s}\right) +\frac{p^{2}}{2m}.
\end{eqnarray}
Notice that in this case, the $\varepsilon _{-}({\bf p})$ branch corresponds
to phonons with a speed of sound given by $u=\sqrt{\frac{\rho V_{s}}{m}}$
that is independent of the interspecies scattering and that $\varepsilon
_{+}({\bf p})$ corresponds to single particle excitations with a mean-field
shift in the energy.

Finally, one may consider the limit that $\rho \alpha \rightarrow -1/2$ and $%
\frac{p^{2}}{2m}\ll \hbar {\cal R}$ which corresponds to the long wavelength
limit in which the two minima are displaced only slightly from $\rho /2.$ In
this case the excitation spectrum has the form 
\end{mathletters}
\begin{equation}
\varepsilon _{\pm }({\bf p})=\sqrt{\Xi _{1}\pm \frac{1}{2}\sqrt{\left(
(\hbar {\cal R})^{2}\left( (2\rho \alpha )^{2}-1\right) -2\rho V_{s}\frac{%
p^{2}}{2m}\right) ^{2}+4(\hbar {\cal R})^{3}\left( (2\rho \alpha
)^{2}-1\right) \frac{p^{2}}{2m}}};
\end{equation}
which for $\rho V_{s}\gg \hbar {\cal R}$, $\varepsilon _{\pm }({\bf p})$
simplifies even further to

\begin{mathletters}
\begin{eqnarray}
\varepsilon _{-}({\bf p}) &=&\sqrt{\frac{\rho \left( V_{x}+V_{s}\right) }{2m}%
}p; \\
\varepsilon _{+}({\bf p}) &=&\sqrt{p^{2}\frac{\rho \left( V_{x}-V_{s}\right) 
}{2m}+\left[ \rho ^{2}\left( V_{x}-V_{s}\right) ^{2}-(\hbar {\cal R})^{2}%
\right] }.
\end{eqnarray}
Again, $\varepsilon _{-}({\bf p})$ corresponds to phonon excitations with a
speed of sound which depends on the average of $V_{s}$ and $V_{x}.$ However,
the $\varepsilon _{+}({\bf p})$ branch can not be given a simple
interpretation in terms of either phonon like collective excitations or
single particle excitations since both the $p^{2}$ term and the term in
brackets are of comparable magnitude.

\section{Discussion}

In this paper we have analyzed the ground state energy of a homogeonous
two-component Bose-Einstein condensate interacting with a spatially uniform
radiation field which couples the two internal states of the condensates. We
have argued that Bose-Einstein condensation occurs in the state which
minimizes $E_{o}$ as a function of the density of one the components (since
the total density is fixed) and the relative phase between the two
components. When the proper ground state for the condensate is chosen, the
quasiparticle excitations above the ground state have real energies and no
instabilities occur. The fact that the quasiparticle spectrum is real and
exhibit no instabilities may not come as a surprise, but these results
differ from those obtained in Ref. \cite{instable}. The instabilities found
in Ref. \cite{instable} originate from an expansion of small amplitude
collective excitations about a point of unstable equilibrium, represented by
the global maximum of Eq. (\ref{ground}).

Throughout this paper it has been assumed that the condensate is in a state
of chemical equilibrium (i.e. $\mu _{a}=\mu _{b}=\mu $ where $\mu _{i}$ is
the chemical potential for component $i$). As such the ground state energy
is found by minimizing $E_{o}$ subject to the constraint $\rho _{a}+\rho
_{b}=\rho .$ An interesting extension of this work would be to consider the
case when the two components are in a state of thermal equilibrium such that
Bose condensation can occur, but not in a state of chemical equilibrium such
as in \cite{griffin}. In this case the population of the two components is
fixed by some external pumping and decay mechanism and the ground state
energy of the condensate will be given by Eq. (\ref{ground}) for {\it fixed} 
$\rho _{a}$ and $\rho _{b}$. In this case one might expect instabilities to
occur in the excitation spectrum since the condensate state is not in
general the minimum of Eq. (\ref{ground}). Normally, these instabilities
would indicate a transition to a phase separated state \cite{phase}.
However, if the pumping and decay are spatially uniform and time independent
so that the relative densities at each point in space are fixed externally,
then it will be impossible for phase separation to occur. The dynamics of
the condensate in this case would be interesting to explore.

\end{mathletters}


\begin{references}
\bibitem{phase}  E. Timmermans, Phys. Rev. Lett. {\bf 81}, 5718 (1998).

\newpage 

Figure 1. Plot of $E_{o}/V$ for $S_{ab}=0$ and $\rho \alpha >0$ showing the
transition from a central maximum at $\rho _{a}=\rho /2$ for $\rho \alpha
<1/2$ to a minimum for $\rho \alpha >1/2$ along with the development of
maxima at $\rho _{a}=\rho ^{(+)},\rho ^{(-)}.$ Note that $E_{o}/V$ is in
units of $\rho \hbar {\cal R}$ and we have subtracted off the constant term $%
\frac{1}{4}(V_{s}+V_{x})\rho ^{2}.$

Figure 2. Plot of $E_{o}/V$ for $S_{ab}=\pi $ and $\rho \alpha <0$ showing
the transition from a central minimum at $\rho _{a}=\rho /2$ for $\rho
\alpha >-1/2$ to a maximum for $\rho \alpha <-1/2$ along with the
development of minima at $\rho _{a}=\rho ^{(+)},\rho ^{(-)}.$ Note that $%
E_{o}/V$ is in units of $\rho \hbar {\cal R}$ and we have subtracted off the
constant term $\frac{1}{4}(V_{s}+V_{x})\rho ^{2}$
\end{references}
\end{document}